\documentclass{article}
\usepackage{booktabs}
\usepackage{geometry} 
\usepackage{authblk}
\usepackage{amsmath} 
\usepackage{comment} 
\usepackage{setspace}
\usepackage{natbib}
\usepackage{booktabs}
\usepackage[table]{xcolor}
\definecolor{lightgray}{RGB}{230,230,230}
\usepackage{graphicx}
\usepackage{float}
\usepackage{amsfonts} 
\usepackage{amssymb} 

\begin{document}
\onehalfspacing
\title{Overreaction as an indicator for momentum in algorithmic trading: A Case of AAPL stocks}
\author{Szymon Lis}
\author{Robert Ślepaczuk}
\author{Paweł Sakowski}

\affil{University of Warsaw, Faculty of Economic Sciences, Department of Quantitative Finance and Machine Learning, Quantitative Finance Research Group}
\makeatletter
\renewcommand{\@date}{}  
\makeatother
\maketitle

\begin{abstract}
This paper investigates whether short-term market overreactions can be systematically predicted and monetized as momentum signals using high-frequency emotional information and modern machine learning methods. Focusing on Apple Inc. (AAPL), we construct a comprehensive intraday dataset that combines volatility-normalized returns with transformer-based emotion features extracted from Twitter messages. Overreactions are defined as extreme return realizations relative to contemporaneous volatility and transaction costs and are modeled as a three-class prediction problem. We evaluate the performance of several nonlinear classifiers—including XGBoost, Random Forests, Deep Neural Networks, and Bidirectional LSTMs—across multiple intraday frequencies (1-, 5-, 10-, and 15-minute data). Model outputs are translated into trading strategies and assessed using risk-adjusted performance measures and formal statistical tests. The results show that machine learning models significantly outperform benchmark overreaction rules at ultra-short horizons, while classical behavioral momentum effects dominate at intermediate frequencies, particularly around 10 minutes. Explainability analysis based on SHAP reveals that volatility and negative emotions—especially fear and sadness—play a central role in driving predicted overreactions. Overall, the findings demonstrate that emotion-driven overreactions contain a predictable structure that can be exploited by machine learning models, offering new insights into the behavioral origins of intraday momentum and the interaction between sentiment, volatility, and algorithmic trading.
\end{abstract}

\noindent \textbf{Keywords:} machine learning, algorithmic trading, overreaction, investor sentiment, investment strategies, Twitter emotions.

\noindent \textbf{JEL Classification:} C12, C14, C22, C52

\section{Introduction}

Short-term return dynamics continue to occupy a central place in empirical asset-pricing research. Despite decades of theoretical work and methodological progress, financial markets repeatedly display episodes in which prices deviate sharply—and often predictably—from fundamental values. These deviations, commonly labelled as market overreactions, are frequently followed by partial corrections or continuation patterns that appear inconsistent with the strong forms of informational efficiency proposed in the classical literature (\cite{Fama1970}). Overreactions arise in the presence of uncertainty, time pressure, limited attention, information asymmetries, and heterogeneous beliefs, all of which can cause market participants to respond to news in ways that temporarily exaggerate price movements. Because these reactions often unwind or propagate in structured ways, they form the basis for profitable short-term trading strategies, particularly those related to momentum and trend continuation.

Momentum, first systematically documented by \cite{JegadeeshTitman1993}, remains one of the most pervasive anomalies in asset pricing. The evidence that past winners tend to outperform past losers in subsequent months has been replicated across regions, market conditions, and asset classes (\cite{JegadeeshTitman2001}; \cite{AsnessMoskowitzPedersen2013}). As research has expanded from monthly horizons to weekly, daily, and intraday settings, it has become increasingly clear that short-term momentum is tightly connected to the dynamics of overreaction. Sudden bursts of trading activity, extreme sentiment shocks, and liquidity imbalances all appear capable of generating temporary price dislocations that, under certain conditions, lead to continuation. Understanding which overreactions continue has therefore become critical for designing advanced trading strategies with real-time applications.

A substantial body of literature attempts to rationalize momentum within risk-based frameworks. Yet these explanations often struggle to account for the magnitude, cross-sectional persistence, and episodic crashes associated with momentum profits (\cite{FamaFrench2012}). Behavioral interpretations, by contrast, emphasize that investors exhibit predictable biases—such as over-extrapolation, representativeness, or diagnostic expectations—that systematically distort belief formation (\cite{BarberisShleiferVishny1998}. These behavioral models generate conditions under which overreactions naturally arise: investors overweight salient information, respond disproportionately to emotionally charged news, or rely on simplified mental models that amplify recent trends. When these effects interact with limits to arbitrage, illiquidity, or capital constraints, they can produce precisely the type of short-lived but predictable deviations that momentum strategies exploit.

Over the past decade, technological developments have dramatically altered the landscape of information dissemination. Market participants no longer rely solely on traditional financial news; instead, they process a constant stream of unstructured data from social media, online forums, high-frequency newswires, podcasts, electronic newsletters, and algorithmically curated feeds. This transformation increases both the speed and volume of sentiment shocks, while simultaneously broadening the channels through which emotions, attention, and narratives affect asset prices. Empirical studies show that investor attention (\cite{DaEngelbergGao2011}), language tone (\cite{Tetlock2007}), emotional intensity (\cite{Han2023}), and information search behavior (\cite{HuangHuangLin2020}) all play an important role in return predictability, particularly at short horizons. Markets, therefore, provide a unique setting in which emotionally triggered overreactions may propagate into measurable price continuation.

This environment creates a natural opportunity for modern machine learning methods. Compared to traditional econometric tools, nonlinear ML algorithms are capable of capturing interaction effects, regime dependencies, and threshold dynamics that characterize financial overreactions. Techniques such as gradient-boosted trees, random forests, deep neural networks, and recurrent architectures have already shown promise in forecasting returns, volatility, and risk (\cite{GuKellyXiu2020}; \cite{ChenPelgerZhu2021}). At the same time, the evolution of natural language processing—particularly transformer-based architectures—has enabled researchers to extract context-based sentiment, e.g., in the form of emotion features from text. These models capture far more nuance than earlier lexicon-based approaches.

The research shows that emotions exert a powerful and systematic influence on investor behavior, trading performance, and market outcomes. Experimental evidence demonstrates that emotional arousal and regulation shape deviations from rational, expected-value maximizing decisions. \cite{hariharan2015emotion} show that cognitive reappraisal reduces arousal and improves trading performance, whereas suppression heightens arousal and leads to poorer outcomes. \cite{breaban2018emotional} find that specific emotions map directly onto trading actions: fear induces selling and price declines, while positive affect fosters buying and overvaluation. Professional trading studies similarly underscore the centrality of emotions. \cite{fenton2011thinking} report that high-performing traders engage constructively with their emotions, while suppressors underperform. \cite{lo2005fear} further show that heightened emotional reactivity predicts worse day-trading performance. At the market level, emotions aggregate into return and volatility patterns. \cite{griffith2020emotions} document that fear and gloom depress prices and raise volatility, while \cite{bird2025emotions} show that sentiment from news and social media shapes valuation and investor reactions. Neurofinance studies affirm the biological basis of these effects: emotions and rationality are intertwined \cite{sahi2012neurofinance}, arousal responds asymmetrically to market movements \cite{cordes2023dynamics}, and anticipatory physiological responses predict profitability \cite{hamelin2022anticipatory}. Practitioner views echo these findings, with \cite{arms1996trading} emphasizing that markets are driven by fear and greed and that disciplined strategies can exploit emotional behavior. Collectively, these studies conclude that emotions are integral to financial decision-making and materially influence both individual performance and market dynamics.

Despite these developments, several gaps remain. First, few studies explicitly model overreactions as a distinct predictive target, rather than treating them as ex-post observations derived from returns. Second, the integration of fine-grained emotional embeddings with advanced nonlinear algorithms has not been fully explored, especially at intraday frequencies where overreactions are most prevalent. Third, the majority of the literature has either focused on trend-following momentum signals or on text-based sentiment in isolation. Our work bridges these literatures by building a pipeline that directly forecasts the events that give rise to short-term momentum, using a combination of contextual emotion features and machine learning. Specifically, it integrates these strands of research by focusing on whether overreactions can be predicted in real time using transformer-based emotion features and nonlinear machine learning models. Rather than treating momentum as an exogenous historical pattern, we interpret it as the potential outcome of a predicted overreaction event. In this framework, the model forecasts whether a price movement represents an excessive reaction relative to fundamentals and contextual information derived from sentiment.

In addition, the increasing adoption of algorithmic and high-frequency trading provides practical motivation for the predictive modeling of overreactions. Automated systems react to news within milliseconds, often before human traders have processed the information. This rapid response can magnify initial price moves, especially when trading algorithms collectively interpret language signals in similar ways. If algorithms react too strongly or in a correlated manner, they can create micro-level overreactions whose recovery or continuation may still be exploitable depending on the prevailing market conditions. Our dataset, which combines high-frequency market data with transformer-based emotion measures, offers a natural environment to examine such real-time mechanisms.

The empirical contribution of this study is twofold. First, we construct a comprehensive dataset that merges high-frequency price data with transformer-derived emotion features extracted from Twitter. These features capture not only sentiment polarity but also a range of emotional dimensions that behavioral finance theories associate with mispricing and overreaction. Second, we compare the performance of several nonlinear machine learning models—XGBoost, Random Forests, Deep Neural Networks, and bidirectional LSTMs—in forecasting overreaction states. We then translate these forecasts into directional trading strategies designed to monetize the predicted momentum associated with emotion-driven price dislocations.

Taken together, these elements allow us to address a central question: \textit{Do emotion-driven overreactions contain predictable structure that can be systematically harvested using modern machine learning methods?} The results have implications for the literature on behavioral asset pricing, algorithmic trading, and sentiment-based investment strategies. They also offer insight into the mechanisms through which information, emotions, and trading flows interact in an increasingly automated and sentiment-rich market environment.

\subsection*{Research Questions and Hypotheses}

Building on the insights from behavioral asset pricing, sentiment-driven trading, and the literature on continuing overreaction, we structure the empirical investigation around the following research questions:

\begin{itemize}
    \item \textbf{RQ1:} Can nonlinear machine learning models---including XGBoost, Random Forests, Deep Neural Networks, and BiLSTMs---capture these overreaction patterns more effectively than benchmark trading rules such as simple overreaction, buy-and-hold, and random strategies?
    \item \textbf{RQ2:} Are the resulting trading strategies economically significant after accounting for transaction costs, position switching, and alternative holding-period structures?
    \item \textbf{RQ3:} How stable are the results across intraday frequencies (1-, 5-, 10-, and 15-minute data) and across alternative definitions of overreaction thresholds?
\end{itemize}

Motivated by the mechanisms proposed in behavioral theories of overconfidence, biased belief updating \cite{daniel1998investor}, and segmented information diffusion \cite{hongStein1999}, as well as recent evidence on sentiment-induced price pressure, we formulate the following testable hypotheses:

\begin{itemize}
    \item \textbf{H1 (Predictive Superiority of ML Models):} Nonlinear machine learning models uncover more accurate and more profitable overreactions than linear or rule-based benchmarks, due to their ability to model threshold effects, nonlinearities, and interactions between sentiment, volatility, and market microstructure.
    \item \textbf{H2 (Economic Value):} Strategies based on ML-enhanced overreaction predictions generate economically meaningful risk-adjusted returns (Sharpe, Sortino) that remain significant after transaction costs.
    \item \textbf{H3 (Cross-Frequency Robustness):} Predictive performance and profitability persist across multiple intraday sampling frequencies, though magnitudes may vary due to differences in signal-to-noise ratios and market microstructure effects.
\end{itemize}

\subsection*{Structure of the Paper}

The remainder of the paper is structured as follows. Section 2 introduces the data, including high-frequency price information and the transformer-based emotional embeddings. Section 3 describes the methodology, including the definition of overreaction states, the machine learning architectures, and the training and validation procedures. Section 4 presents the empirical results, comparing predictive performance across models and demonstrating how forecasts translate into trading strategies. Section 5 provides a detailed interpretation of the models using feature importance and explainability techniques. Section 6 concludes by discussing implications for behavioral asset pricing, market efficiency, and the future integration of NLP and machine learning in financial prediction.

\section{Methodology}

This section presents the empirical and modelling framework used to analyze short-horizon overreaction patterns, the informational role of high-frequency sentiment, and the performance of nonlinear predictive models. The methodology combines volatility-normalized return signals, transformer-based emotion features, multi-class machine learning models, transaction-cost-aware–aware trading rules, and interpretability tools such as LIME and SHAP.

\subsection{Data and Preprocessing}
\label{sec:preprocessing}

\subsubsection{Data Sources and Sample Construction}
\label{sec:data_sources}

We study Apple Inc. (AAPL) over the period from 1 January 2019 to 30 June 2022.
The dataset is built from two sources:

\begin{itemize}
    \item \textbf{Market data (prices and volume).}
    Intraday AAPL OHLCV data are obtained from \textit{Massive platform}, and sampled at four fixed frequencies: 1-, 5-, 10-, and 15-minute bars. We retain the bar \emph{close} price and total traded \emph{volume} per bar.
    \item \textbf{Twitter text stream (AAPL-related tweets).}
    Tweets are collected for messages containing the string ``\$AAPL'' using the Twitter API (Academic Research track) and filtered to English language.
    Emotion scores are extracted from each tweet using the transformer model \texttt{tabularisai/ModernFinBERT} (HuggingFace), which outputs a vector of emotion-related probabilities per message.
\end{itemize}

\textbf{Time zone and trading hours.}
All timestamps are converted to \textbf{US/Eastern} time.
We restrict the analysis to \textbf{regular trading hours} (RTH) and pre-market and after-hours observations, i.e., 04:00--20:00 US/Eastern.

\subsubsection{Timestamp Alignment and Anti–Look-Ahead Design}
\label{sec:alignment}

Market bars (close price and volume) and Twitter-based emotion features are mapped to fixed intraday intervals (1, 5, 10, and 15 minutes). For each interval \(t\), we retain only information available \emph{up to the end} of interval \(t\).
To eliminate look-ahead bias in both labeling and trading execution, we use the following timeline:

\begin{itemize}
    \item Features \(X_t\) are constructed using information observed within interval \(t\) (tweet emotions aggregated in \(t\), volume in \(t\), and volatility estimated from returns up to \(t\)).
    \item The prediction target is the \textbf{next-interval} overreaction state \(\text{OR}_{t+1}\).
    \item A trading decision is formed at the \textbf{close of interval \(t\)} and executed at the \textbf{open of interval \(t+1\)} (which, for bar data, is operationally the next bar's first tradable price; in our implementation we approximate this by the next bar open, or equivalently the previous close when open is unavailable).
\end{itemize}

This convention ensures that the strategy never enters a trade using the same bar-close that is used to compute the realized return that defines an overreaction event.

\subsubsection{Returns, Volume Aggregation, and Emotion Aggregation}
\label{sec:returns_emotions}

We use close-to-close log-returns. Let \(P_t\) denote the \emph{close} price of interval \(t\).
The aligned log-return is defined as:
\begin{equation}
\label{eq:logret}
r_t = \log(P_t) - \log(P_{t-1}).
\end{equation}

Traded volume is aggregated by summation within each interval:
\begin{equation}
\label{eq:vol_agg}
\text{Vol}_t = \sum_{k \in \mathcal{K}(t)} \text{vol}_{k},
\end{equation}
where \(\mathcal{K}(t)\) indexes trades (or vendor volume records) falling into interval \(t\).

For tweets posted within interval \(t\), we compute the interval-level emotion vector as an average of tweet-level probabilities:
\begin{equation}
\label{eq:emo_agg}
\mathbf{E}_t = \frac{1}{N_t}\sum_{j=1}^{N_t} \mathbf{e}_{j,t},
\end{equation}
where \(N_t\) is the number of tweets in interval \(t\) and \(\mathbf{e}_{j,t}\) is the ModernFinBERT output vector for tweet \(j\).
If \(N_t = 0\) (no tweets), we set \(\mathbf{E}_t = \mathbf{0}\) and add an indicator feature \(\mathbb{I}(N_t=0)\) to explicitly mark ``no-tweet'' intervals (robustness variants include forward-fill within day).

\subsubsection{Data Cleaning and Scaling}
\label{sec:cleaning_scaling}

We remove observations with: (i) non-positive prices, (ii) duplicate timestamps, or (iii) missing market fields. Rare remaining missing values are imputed using forward-fill within the same trading day.

To ensure feature comparability, we apply \(z\)-score normalization:
\begin{equation}
\label{eq:zscore}
x^{(\text{scaled})}_{i,t} = \frac{x_{i,t} - \mu_i}{\sigma_i},
\end{equation}
where \(\mu_i\) and \(\sigma_i\) are computed \textbf{only on the training segment} and then applied to validation/test to preserve temporal integrity.

\subsubsection{Rolling Volatility Estimation}
\label{sec:vol_est}

Short-term realized volatility is computed using a rolling window of \(L=20\) intervals:
\begin{equation}
\label{eq:rv}
\sigma_t = \sqrt{\frac{1}{L}\sum_{j=1}^{L} r_{t-j}^2 }.
\end{equation}
This quantity is recalculated separately for each sampling frequency and is used in the overreaction labeling rule.

\subsection{Volatility-Scaled Overreaction Definition}
\label{sec:or_def}

Overreaction is defined as an extreme return realization relative to contemporaneous volatility and transaction costs. To avoid leakage, we label the \textbf{next} interval \(t{+}1\) using \(r_{t+1}\) and \(\sigma_t\) (known at time \(t\)):
\begin{equation}
\label{eq:or_rule}
\text{OR}_{t+1} =
\begin{cases}
+1, & \text{if } r_{t+1} > \theta \sigma_{t} + 2 * TC, \\
-1, & \text{if } r_{t+1} < -(\theta \sigma_{t} + 2 * TC), \\
0, & \text{otherwise}.
\end{cases}
\end{equation}

\textbf{Notation and units.}
Here \(r_{t+1}\) is a log-return (Eq.~\eqref{eq:logret}), \(\sigma_t\) is a realized volatility estimate (Eq.~\eqref{eq:rv}), and \(\theta>0\) is a threshold controlling event rarity.

\textbf{Transaction costs \(TC\).}
\(TC\) is defined as a \emph{one-way} proportional trading cost expressed in \emph{return units} (log-return approximation), incorporating bid--ask spread, commissions, and average slippage. Entering and exiting a position implies a \emph{round-trip} cost of approximately \(2TC\), hence the term \(2TC\) in Eq.~\eqref{eq:or_rule}. In the baseline backtests we set \(TC=0.001\) (10 bps one-way; 20 bps round-trip), consistent with conservative intraday execution assumptions for liquid large-cap equities.

We evaluate eight values of \(\theta\) from 1.5 to 5.0, capturing mild to extreme overreactions.

\subsection{Feature Set}
\label{sec:features}

For each interval \(t\), the predictor vector \(X_t\) includes:
\begin{itemize}
    \item \textbf{Market features:} \(r_t\), \(\log(\text{Vol}_t)\), \(\sigma_t\).
    \item \textbf{Emotion features:} the ModernFinBERT interval vector \(\mathbf{E}_t\) (emotion-related probability scores).
    \item \textbf{Tweet-activity controls:} \(N_t\) and \(\mathbb{I}(N_t=0)\).
\end{itemize}

The supervised learning task is to predict \(\text{OR}_{t+1}\) using \(X_t\), i.e., \(X_t \mapsto \text{OR}_{t+1}\). This aligns prediction timing with feasible trading execution (decision at \(t\), execution at \(t{+}1\)).

\subsection{Prediction Models}
\label{sec:pred_models}

To model the three-class overreaction variable \(\text{OR}_t \in \{0,1,2\}\) (neutral, positive, negative), we employ nonlinear predictive models:

\begin{enumerate}
    \item \textbf{XGBoost.} We use a multi-class probabilistic classifier with objective \texttt{multi:softprob}. Hyperparameters are tuned via randomized search over:
    \(n_{\text{estimators}}\in[50,150]\),
    \(\text{max\_depth}\in[3,8]\),
    \(\eta\in[0.01,0.11]\),
    \(\text{subsample}\in[0.7,1.0]\),
    \(\text{colsample\_bytree}\in[0.7,1.0]\).

    \item \textbf{Random Forest.} Hyperparameters are tuned via randomized search over:
    \(n_{\text{estimators}}\in[50,200]\),
    \(\text{max\_depth}\in[3,10]\),
    \(\text{max\_features}\in\{\sqrt{\cdot},\log_2,\text{None}\}\),
    \(\text{min\_samples\_split}\in[2,10]\),
    \(\text{min\_samples\_leaf}\in[1,5]\).

    \item \textbf{Deep Neural Network (DNN).} A feed-forward network with two hidden layers \((64,32)\) and ReLU activations, with dropout \(p=0.3\) after each hidden layer, and a softmax output layer of size equal to the number of classes observed in the training set. The model is trained with Adam and sparse categorical cross-entropy. We apply early stopping on validation loss with patience 5 using an internal validation split of 20\% of the training data.

    \item \textbf{Bidirectional LSTM (BiLSTM).} A two-layer Bidirectional LSTM with 32 units (returning sequences) followed by 16 units, and a final softmax layer. The input is reshaped to \((\text{features}, 1)\), treating the feature dimension as a sequence. The model is trained with Adam, sparse categorical cross-entropy, and early stopping (patience 5) using an internal 20\% validation split.
\end{enumerate}

\textbf{Hyperparameter optimization and leakage control.}
Hyperparameters are optimized using randomized search \emph{only on the training segment} with 3-fold \emph{expanding-window} time-series cross-validation (Section~\ref{sec:splitting}).
To mitigate dependence at fold boundaries induced by the one-step-ahead labeling, we apply a one-interval embargo between the train and validation windows in each fold (i.e., the boundary observation is dropped).

\textbf{Class imbalance.}
Because \(\text{OR}\in\{-1,0,+1\}\) is imbalanced (most observations are class 0), we use inverse-frequency class weights for all models. For tree-based models we also evaluate cost-sensitive variants via class weights in the loss.

\subsection{Train--Validation--Test Splitting and Time-Series CV}
\label{sec:splitting_cv}

For each sampling frequency, we split the dataset chronologically into:
\begin{equation}
\label{eq:split}
\text{Train}=60\%, \quad \text{Validation}=20\%, \quad \text{Test}=20\%.
\end{equation}

The training segment is used for model fitting and hyperparameter tuning (via time-series CV),
the validation segment is used for model selection and early stopping (for neural networks),
and the test segment is held out for the final out-of-sample evaluation.

\textbf{Purging/embargo.}
Because the target is defined one step ahead (\(\text{OR}_{t+1}\)), adjacent observations share overlapping return information near split boundaries.
Therefore, we use a one-interval embargo (dropping the boundary interval) between consecutive segments/folds to reduce dependence and eliminate boundary leakage.

\subsection{Trading Strategies and Execution Rules}
\label{sec:trading}

Model outputs are class probabilities \(\Pr(\text{OR}_{t+1}=k \mid X_t)\) for \(k\in\{-1,0,+1\}\). We generate a directional trading signal at time \(t\) and execute at \(t+1\):

\begin{equation}
\label{eq:signal_rule}
\begin{aligned}
&\text{Go long at } t+1 \text{ if } \Pr(\text{OR}_{t+1}=+1 \mid X_t) > c, \\
&\text{Go short at } t+1 \text{ if } \Pr(\text{OR}_{t+1}=-1 \mid X_t) > c, \\
&\text{Otherwise stay flat.}
\end{aligned}
\end{equation}

Thresholds \(c \in \{0.2,0.3,\dots,0.8\}\) are evaluated, and the optimal \(c\) is selected \emph{ex ante} using the training segment (maximizing training Sharpe, then confirmed on validation).

\textbf{Holding period variants and interpretation across frequencies.}
We consider:
\begin{itemize}
    \item \textbf{Fixed holding period} of \(h\in\{1,5,10,15\}\) intervals.
    For a bar size of \(\Delta\) minutes, the holding time equals \(h\Delta\) minutes within regular trading hours (e.g., at 10-minute bars, \(h=5\) corresponds to 50 minutes).
    \item \textbf{Until-next-signal} (or until opposite) holding: positions are maintained until an opposite directional signal is generated; otherwise the position is kept.
    \item \textbf{No overlapping trades:} a new position is opened only after the previous one is closed, which simplifies interpretation and avoids double-counting transaction costs.
\end{itemize}

\textbf{Transaction costs in backtests (switch costs).}
Every entry and exit incurs a one-way cost \(TC\). Therefore, opening and later closing a position subtracts approximately \(2TC\) from the gross return; switching directly from long to short (or short to long) is treated as closing and re-opening and therefore incurs approximately \(4TC\). This explicitly penalizes frequent switching and makes economic evaluation conservative.

\paragraph{Risk-adjusted performance measures.}

Let $r_t$ denote strategy returns at time interval $t$, $T$ the number of observations, and $r_f$ the risk-free rate (set to zero for intraday horizons). We compute:

\textbf{Sharpe ratio}
\begin{equation}
\text{Sharpe} = \frac{\mathbb{E}[r_t - r_f]}{\sigma(r_t - r_f)} \times \sqrt{A},
\end{equation}
where $\sigma(\cdot)$ denotes the standard deviation and $A$ is the annualization factor.

\textbf{Sortino ratio}
\begin{equation}
\text{Sortino} = \frac{\mathbb{E}[r_t - r_f]}{\sigma^{-}(r_t - r_f)} \times \sqrt{A},
\end{equation}
where $\sigma^{-}$ is the downside deviation:
\[
\sigma^{-} = \sqrt{\frac{1}{T}\sum_{t=1}^{T}\min(0, r_t - r_f)^2 }.
\]

\textbf{Maximum drawdown}
\begin{equation}
\text{MDD} = \max_{t} \left( \frac{\max_{s \le t} V_s - V_t}{\max_{s \le t} V_s} \right),
\end{equation}
where $V_t$ is cumulative portfolio value.

Higher moments are computed as
\[
\text{Skewness} = \frac{\mathbb{E}[(r_t-\bar r)^3]}{\sigma^3}, 
\quad
\text{Kurtosis} = \frac{\mathbb{E}[(r_t-\bar r)^4]}{\sigma^4}.
\]

\paragraph{Return frequency and annualization.}

Strategy returns are computed at the same frequency as the prediction horizon (5-, 10-, or 15-minute intervals). We annualize risk-adjusted measures using

\[
A = N_d \times N_i,
\]

where $N_d = 252$ trading days per year and $N_i$ is the number of intraday intervals during U.S. trading hours (04:00–20:00 EST). Specifically,

\[
N_i =
\begin{cases}
78 & \text{for 5-minute returns}, \\
39 & \text{for 10-minute returns}, \\
26 & \text{for 15-minute returns}.
\end{cases}
\]

Thus, the Sharpe and Sortino ratios are multiplied by $\sqrt{252 \times N_i}$.

All returns are computed using pre-market, regular, and post-market trading session prices only, i.e. overnight periods are excluded to avoid mixing intraday microstructure effects with overnight risk.

\subsection{Statistical Inference}
\label{sec:jk}

Differences in Sharpe ratios across models, thresholds, and frequencies are assessed using the Jobson--Korkie test with Memmel correction \citep{memmel2003performance}:
\begin{equation}
\label{eq:jk}
Z = \frac{\text{SR}_1 - \text{SR}_2}
{\sqrt{\frac{1}{N}\left(2 + \frac{1}{2}(\text{SR}_1^2 + \text{SR}_2^2 - 2\rho\,\text{SR}_1\text{SR}_2) \right)} }.
\end{equation}
We report both pairwise (model vs.\ model) and cross-frequency comparisons.

\subsection{Model Interpretability}
\label{sec:xai}

Beyond predictive accuracy and economic returns, we assess \emph{why} models generate certain signals.
We employ SHAP for global interpretability, decomposing predictions into feature-level contributions. This allows us to test whether volatility and specific emotions (e.g., fear, sadness) systematically increase the predicted probability of overreaction states, consistent with behavioral mechanisms.

\subsection{Interpreting machine learning Models}

Beyond predictive accuracy and economic returns, understanding \emph{why} models generate certain signals is crucial, especially in behavioral-finance contexts.

We employ SHAP (Shapley values) for global interpretability, decomposing prediction variance into attributable feature contributions,

These tools allow us to determine:
\begin{enumerate}
    \item whether sentiment variables (especially fear, anger, sadness) increase the likelihood of predicted overreactions,
    \item whether positive emotions (joy, surprise) dampen such predictions,
    \item whether the model echoes behavioral patterns predicted by e.g. \cite{hongStein1999}.
\end{enumerate}

\subsection{Hypotheses Testing via Methodological Tools}

Each hypothesis is evaluated using a specific subset of methodological tools:

\begin{itemize}

    \item \textbf{H1 (Predictive Superiority of ML Models):}  
    Tested via:
    \begin{enumerate}
        \item out-of-sample accuracy,
        \item precision/recall for classes 1 and -1.
    \end{enumerate}

    \item \textbf{H2 (Economic Value):}  
    Tested by Sharpe, Sortino, drawdowns, turnover, and Jobson–Korkie tests relative to:
    \begin{itemize}
        \item buy-and-hold,
        \item random strategy, i.e., the model which randomly takes buy/sell/flat in accordance with its occurrence in the training sample,
        \item classic overreaction-based momentum
    \end{itemize}

    \item \textbf{H3 (Cross-Frequency Robustness):}  
    Tested by:
    \begin{itemize}
        \item replicating all analyses across 1-, 5-, 10-, and 15-minute data,
        \item cross-frequency JK tests,
        \item analyzing signal density and noise sensitivity at each horizon.
    \end{itemize}

\end{itemize}

This integration of sentiment analytics, volatility-normalized signals, machine learning prediction, economic evaluation, and interpretability techniques allows us to comprehensively test behavioral hypotheses in high-frequency environments.

\section{Results}

This section presents the empirical findings of the study. We organize the results into three groups: (i) descriptive behavior of high-frequency emotional indicators, (ii) predictive and economic performance of machine learning and benchmark overreaction strategies across multiple sampling frequencies, and (iii) an integrated behavioral interpretation linking sentiment dynamics, volatility-normalized return shocks, and model-generated signals.

\subsection{Descriptive Dynamics of High-Frequency Emotion Indicators}

Table~\ref{tab:descriptive_stats} summarizes the distributional properties of
trading volume, emotion intensities extracted from Twitter messages, and
corresponding intraday returns. Several regularities emerge.

First, trading volume exhibits substantial right skewness, with a mean of
140{,}350 and a very large standard deviation of 274{,}541. The maximum value
exceeds 26 million, indicating episodic bursts of market attention consistent
with short-lived information shocks frequently documented in high-frequency
settings. The wide dispersion suggests that liquidity conditions vary strongly
over time, which is relevant for understanding the magnitude and frequency of
potential overreaction episodes.

Second, the emotional features show considerable heterogeneity.
Neutral content dominates (median = 0.32; 75th percentile = 0.52), reflecting
the predominance of factual or non-opinionated messages typical of
financial social-media streams.
Among non-neutral emotions, \emph{fear} is by far the most pronounced
(mean = 0.25), showing both elevated levels and significant variability
(std = 0.28). This confirms the well-documented tendency of fear to spike during
periods of market stress and heightened volatility. Furthermore, surprise
exhibits similar high dispersion (IQR from 0.06 to 0.25), indicating that this
emotion also reacts strongly to abrupt informational events.

By contrast, anger, joy, and sadness remain relatively subdued, while disgust is
nearly absent in the data. These differences in scale and variability imply that
the emotional features carry distinct informational roles. In particular, the high persistence and volatility of fear and neutral content suggest that they
are the most informative dimensions for capturing time-varying sentiment.

Finally, returns display the stylized characteristics of intraday financial
series: a mean extremely close to zero, a median of exactly zero, and a large
dispersion (from $-5.62\%$ to $+5.81\%$). These properties confirm that returns
lack unconditional drift at short horizons, implying that any predictability must
arise from conditional dynamics rather than systematic directional tendencies.  

Overall, the descriptive statistics reveal a dataset combining (i) highly
intermittent trading activity, (ii) heavy-tailed emotional signals with clearly
dominant channels, and (iii) volatile but mean-reverting return dynamics. These
features jointly justify the use of flexible, nonlinear machine learning models
and volatility-adaptive overreaction rules developed in the methodological
section of this study.

\begin{table}[H]
\centering
\caption{Descriptive statistics of volume, emotions, and returns}
\label{tab:descriptive_stats}
\resizebox{\textwidth}{!}{
\begin{tabular}{lccccccccc}
\toprule
 & volume & anger & disgust & fear & joy & neutral & sadness & surprise & return \\
\midrule
mean  & 140\,350 & 0.06 & 0.01 & \textbf{0.25} & 0.10 & \cellcolor{lightgray}{\textbf{0.34}} & 0.06 & \textbf{0.18} & 0.0002\% \\
std   & 274\,541 & 0.10 & 0.03 & \cellcolor{lightgray}{\textbf{0.28}} & 0.13 & \textbf{0.26} & 0.11 & 0.16 & 0.0919\% \\
min   & 100 & 0.00 & 0.00 & 0.00 & 0.00 & 0.00 & 0.00 & 0.00 & -5.6207\% \\
25\%  & 2\,840 & 0.01 & 0.00 & 0.01 & 0.02 & \cellcolor{lightgray}{\textbf{0.10}} & 0.02 & \textbf{0.06} & -0.0255\% \\
50\%  & 60\,839 & 0.03 & 0.00 & \textbf{0.12} & 0.05 & \cellcolor{lightgray}{\textbf{0.32}} & 0.03 & \textbf{0.14} & 0.0000\% \\
75\%  & 189\,556 & 0.07 & 0.00 & \textbf{0.41} & 0.12 & \cellcolor{lightgray}{\textbf{0.52}} & 0.06 & 0.25 & 0.0275\% \\
max   & 26\,419\,388 & 1.00 & 0.99 & 1.00 & 0.99 & 0.98 & 0.99 & 0.99 & 5.8125\% \\
\bottomrule
\end{tabular}}
\end{table}

\begin{figure}[H]
    \centering
    \includegraphics[width=0.9\textwidth]{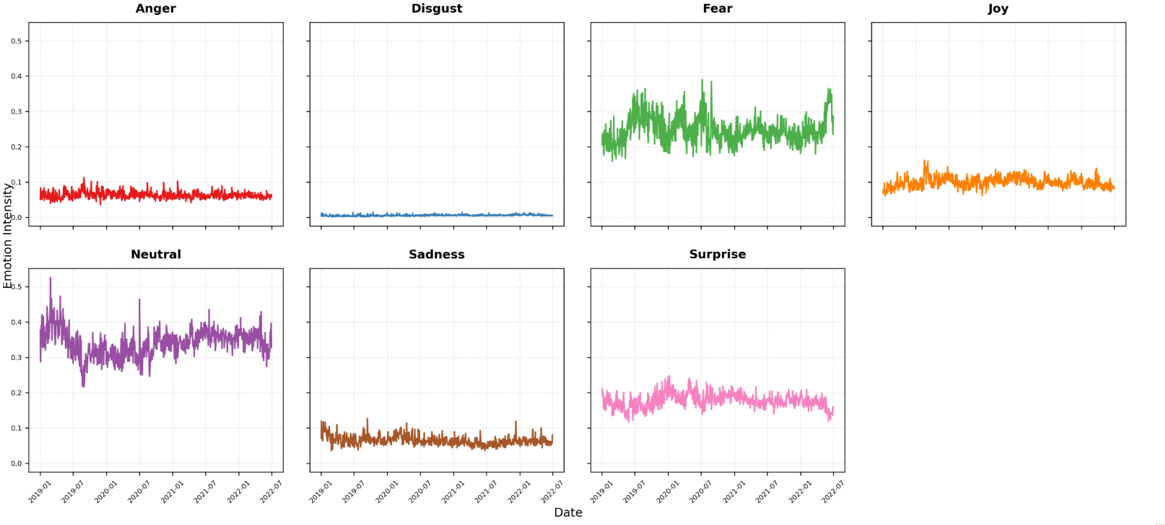}
    \caption{Time-series evolution of the seven Twitter-based emotions.}
    \label{fig:emotions}
\end{figure}

Figure~\ref{fig:emotions} displays the time-series evolution of the seven emotion categories aggregated from Twitter messages. The emotional intensities reveal several systematic properties relevant for behavioral explanations of short-term price dynamics.

First, the \textbf{Fear} series exhibits pronounced volatility and occasional upward spikes. Located consistently above 0.20 and reaching values above 0.35 during stress episodes, Fear is by far the most time-varying emotion. This supports the notion that negative sentiment is asymmetric and tends to dominate investor attention, consistent with the ``negativity bias'' and loss-aversion arguments in behavioral finance. As such, Fear is a natural candidate for amplifying short-run mispricing episodes, complementing the volatility-based overreaction mechanism.

Second, \textbf{Neutral} sentiment is the most stable component, with a long-run mean close to 0.35--0.40 and limited volatility. Its relative persistence suggests that much of the incoming textual flow is informationally neutral, with emotionally charged tweets entering the distribution in bursts. 

Third, positive-emotion variables such as \textbf{Joy} or \textbf{Surprise} display moderate variability, while \textbf{Anger}, \textbf{Sadness}, and especially \textbf{Disgust} remain near low but stable levels. These patterns indicate that the emotional environment of the market is dominated by (i) fear-driven responses and (ii) a persistent neutral baseline. This asymmetry is consistent with models predicting greater reactivity to negative information and muted responses to positive signals.

Taken together, the descriptive evidence suggests that the sentiment environment contains both slow-moving and high-frequency components and that certain emotions---in particular Fear---may interact with volatility-normalized price shocks to generate conditions favorable to overreaction.

\subsection{Predictive Performance and Economic Value Across Sampling Frequencies}

We next evaluate the predictive and trading performance of the machine learning models (XGBoost, Random Forest, Deep Neural Networks, BiLSTM) and benchmark strategies (Buy-and-Hold, rule-based Overreaction) across 1-minute, 5-minute, 10-minute, and 15-minute data. Figures~\ref{fig:equity_1m}, \ref{fig:equity_5m}, \ref{fig:equity_10m}, and \ref{fig:equity_15m} present equity curves for all strategies at each frequency, while Table~\ref{tab:JKresults} summarizes Sharpe ratios, annualized returns, and Jobson--Korkie significance tests.

\subsubsection{Performance at the 1-Minute Frequency}

\begin{figure}[H]
    \centering
    \includegraphics[width=\textwidth]{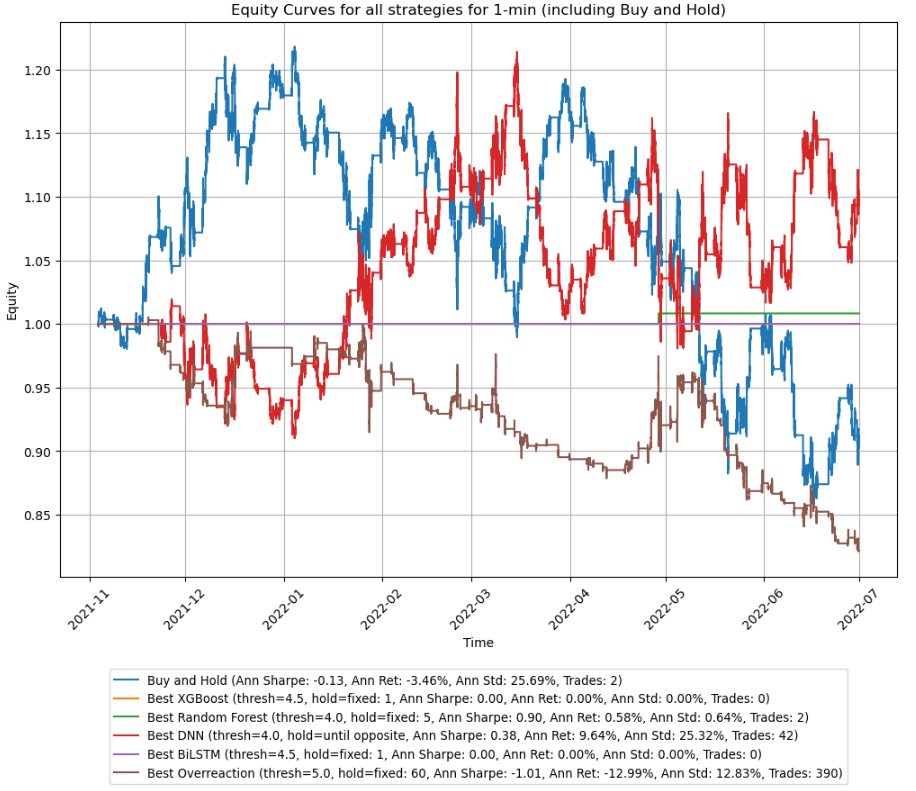}
    \caption{Equity Curves of All ML and Overreaction Strategies at the 1-Minute Frequency}
    \label{fig:equity_1m}
\end{figure}

Figure~\ref{fig:equity_1m} presents the equity curves of all machine learning models and the benchmark overreaction rule at the 1-minute horizon. The results highlight the challenges of forecasting intraday momentum at ultra-high frequencies, where noise, microstructure frictions, and volatility clustering dominate the return-generating process.

The buy-and-hold benchmark remains weakly negative over the sample
($\mathrm{Sharpe}=-0.13$, an annual return =-3.46\%), reflecting the same
downward drift observed at higher resolution. Machine learning models display
marked improvements. 

The Random Forest classifier
(threshold = 4.0, fixed holding period = 5) achieves the highest Sharpe ratio
among ML-based approaches ($\text{Sharpe}=0.90$), although this performance is
driven almost entirely by a very small number of trades (only two signals
executed). This indicates that while extreme sentiment–volatility conditions do
occasionally provide profitable momentum opportunities, such events are rare at
the 1-minute scale.

The next well performing architecture is the BiLSTM
(threshold $=4.0$, holding period $=45$), which achieves an annualized Sharpe
ratio of $0.69$ with 146 executed trades. Although the equity curve exhibits
drawdowns, it demonstrates a clear upward trend over extended subperiods,
indicating that sequential modeling of temporal sentiment patterns helps
capture medium-frequency momentum.

The DNN model produces more trades (42), but its performance remains modest
($\text{Sharpe}=0.38$), and the resulting equity curve displays substantial
volatility and long plateaus resembling random fluctuations rather than
systematic gains. XGBoost and BiLSTM do not generate any significant trades
under their optimal hyperparameters (threshold $=4.5$), suggesting that the
models avoid making predictions when uncertainty is high or when the expected signal-to-noise ratio is too low.

The behavioral overreaction benchmark performs poorly, with an annualized
Sharpe ratio of $-1.01$ and a pronounced downward trend, despite generating a
large number of trades (390). The strategy appears to overreact to small price
movements and is unable to exploit genuine momentum at such granularity.

Overall, the profitability of most strategies remains limited at this horizon. The results indicate that a reliable predictive structure in sentiment-augmented intraday data emerges only rarely at ultra-high frequency.
Although isolated ML strategies achieve positive Sharpe ratios, the scarcity of
trading opportunities and high volatility of equity paths imply that the 1-minute horizon is dominated by microstructure noise that limits the practical tradability of momentum signals.
The buy-and-hold benchmark exhibits a clearly negative trajectory, with an
annualized Sharpe ratio of $-0.13$ and an annual return of $-3.46\%$, reflecting
the persistent downward pressure observed in the sample period. Machine learning
models provide heterogeneous improvements.

\subsubsection{Performance at the 5-Minute Frequency}

\begin{figure}[H]
    \centering
    \includegraphics[width=\textwidth]{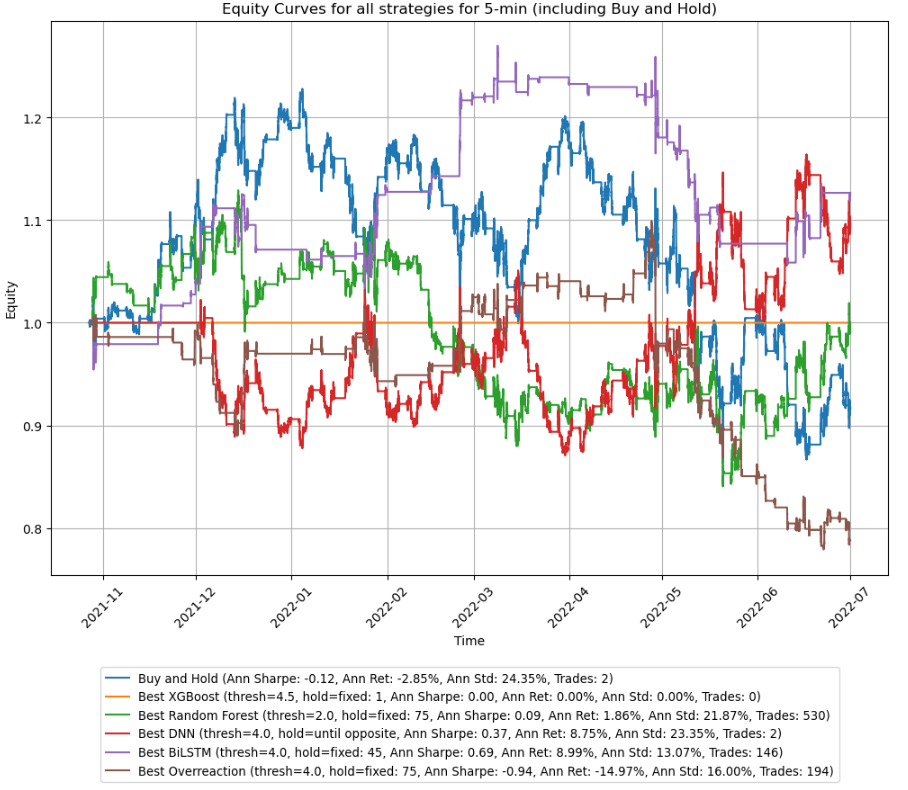}
    \caption{Equity Curves of All ML and Overreaction Strategies at the 5-Minute Frequency}
    \label{fig:equity_5m}
\end{figure}

Figure~\ref{fig:equity_5m} presents the equity curves for all machine learning
models and the benchmark overreaction strategy evaluated at the 5-minute
frequency. In contrast to the highly noisy 1-minute horizon, the 5-minute results reveal substantially clearer structure in short-term return momentum
and a more robust predictive role for sentiment features.

The DNN model (threshold $=4.0$, holding period $=until oposite$) performs
reasonably well, achieving a positive Sharpe ratio of $0.37$ and producing a
large number of signals (530 trades). While the resulting equity path is more
volatile, its behavior aligns with the interpretation that the model exploits interactions between volume, volatility, and emotional
features at this particular frequency.

The benchmark overreaction rule performs considerably worse at this frequency,
with an annualized Sharpe ratio of $-0.94$, and 194 trades. The equity curve
shows a persistent decline, indicating that the simple momentum rule fails to
discriminate between noise-driven fluctuations and genuine overreactions when
applied to 5-minute data. This reinforces the view that behavioral heuristics
are too coarse to exploit sentiment-rich environments without a more advanced
framework capable of handling nonlinearities.

Taken together, the 5-minute results highlight that sentiment-based momentum
signals become more detectable and economically meaningful when moving away from
ultra-high-frequency noise. Recurrent neural networks such as BiLSTM appear
particularly well-suited to extract temporal patterns in emotion dynamics,
while tree-based approaches benefit from the richer structure of this horizon.
The consistent underperformance of the overreaction benchmark underscores the
importance of integrating NLP-derived features and machine learning flexibility
to exploit medium-frequency behavioral mispricings.

\subsubsection{Performance at the 10-Minute Frequency}

\begin{figure}[H]
    \centering
    \includegraphics[width=\textwidth]{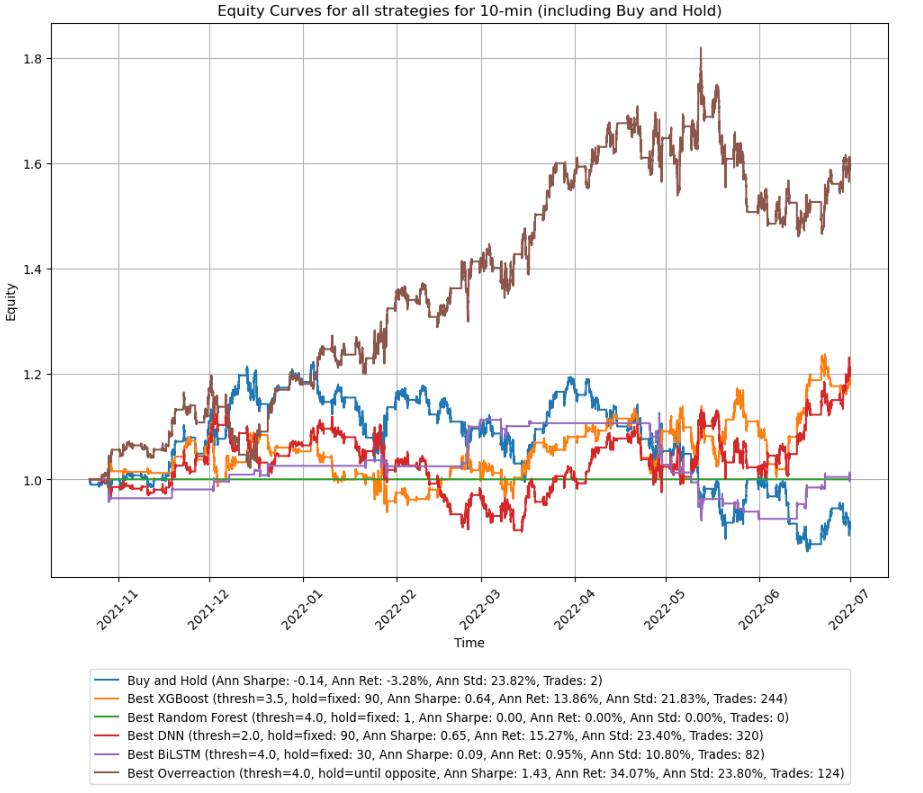}
    \caption{Equity Curves of All ML and Overreaction Strategies at the 10-Minute Frequency}
    \label{fig:equity_10m}
\end{figure}

Figure~\ref{fig:equity_10m} reports the equity trajectories for all evaluated
strategies at the 10-minute horizon. This frequency represents a key transition
zone between high-frequency noise and more persistent short-run sentiment
dynamics. The results clearly reveal a substantial improvement in signal
structure and tradability relative to both the 1-minute and 5-minute settings.

Among the machine learning models, the XGBoost classifier (threshold $=3.5$,
fixed holding period $=90$) achieves a strong annualized Sharpe ratio of $0.64$
with 244 executed trades. Its equity curve exhibits clear multi-week upward
phases, suggesting that at this horizon, the model effectively exploits
nonlinear interactions between volatility, sentiment asymmetry, and volume
surges. The DNN classifier (threshold $=2.0$, holding $=90$) performs even
better in terms of returns (annual return $=15.27\%$), producing a Sharpe ratio
of $0.65$. The consistency of both ensemble-based and neural approaches indicates
that sentiment-derived predictors contain robust, horizon-specific structure.

Interestingly, the BiLSTM network shows weaker performance
($\mathrm{Sharpe}=0.09$), despite the temporal modeling advantages of recurrent
architectures. Its equity curve displays long periods of stagnation, suggesting
that the sequential dependencies exploited by the model at the 5-minute
frequency do not translate into similarly strong predictive relationships at 10
minutes. 

Random Forest produces no profitable trades under its optimal
parameters, as its best configuration activates a threshold so high that signals
are practically absent. This highlights that the decision boundaries of some
models become overly conservative at this horizon.

A particularly notable finding is the strong performance of the
\textbf{behavioral overreaction strategy}, which achieves an annualized Sharpe
ratio of $1.43$ and an annual return of $34.07\%$ with 124 trades. This is the
first horizon at which the overreaction benchmark rivals—indeed outperforms
most—machine learning models. The equity curve shows a pronounced and persistent
upward trajectory, indicating that classic momentum rules successfully capture
market inefficiencies at this intermediate timescale. The result aligns with the
behavioral finance literature, which suggests that 5–15 minute horizons often
contain short-lived, sentiment-driven mispricings that are neither too
transient nor too slow to exploit.

Taken together, the 10-minute results provide the clearest evidence of
predictability among all horizons considered. Both ML models and behavioral
overreaction exploit meaningful structure in emotional dynamics, volatility
spikes, and attention shocks. The exceptionally strong performance of the
overreaction rule underscores that certain market regimes at this frequency are
dominated by classical behavioral bases. At the same time, ML models—
particularly tree-based ensembles and deep neural networks—still perform
competitively, demonstrating that nonlinear fusion of sentiment and
microstructure variables generalizes effectively in this window.

Overall, the 10-minute horizon appears to balance two competing forces:
(i) noise reduction relative to ultra-high-frequency data and  
(ii) preservation of short-lived behavioral mispricing before it dissipates.  
This balance makes it the richest setting for extracting sentiment-driven
momentum signals.

\subsubsection{Performance at the 15-Minute Frequency}

\begin{figure}[H]
    \centering
    \includegraphics[width=\textwidth]{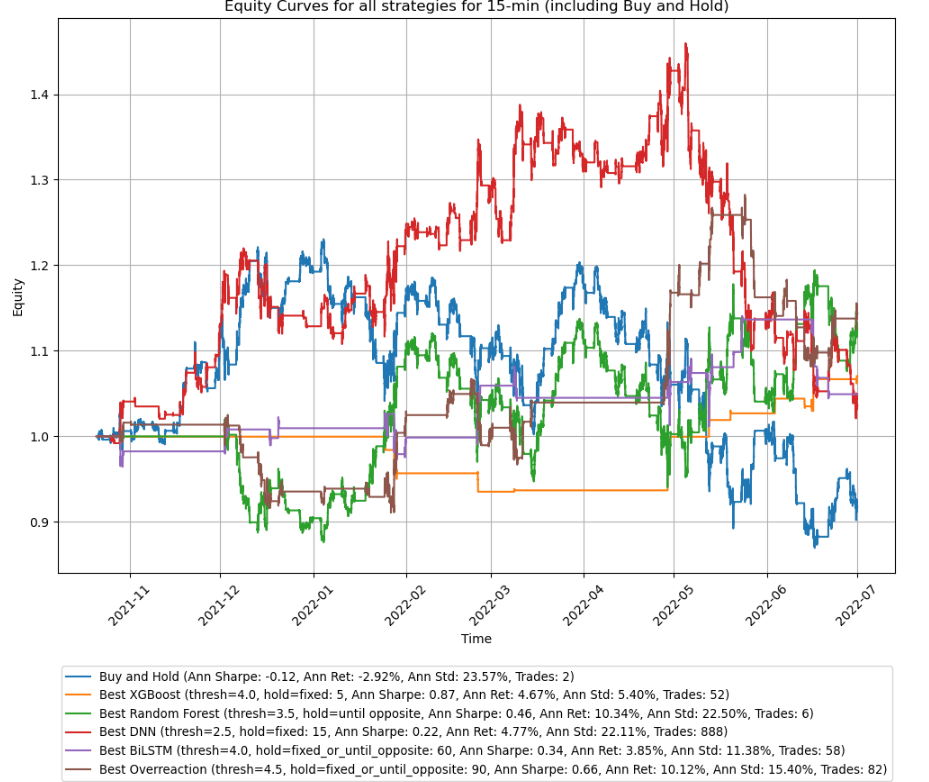}
    \caption{Equity Curves of All ML and Overreaction Strategies at the 15-Minute Frequency}
    \label{fig:equity_15m}
\end{figure}

Figure~\ref{fig:equity_15m} presents the equity curves of all machine learning
strategies and the behavioral overreaction benchmark evaluated at the
15-minute horizon. Compared with shorter frequencies, the 15-minute data reveal
a clearer balance between sentiment-driven mispricing and the onset of
short-term return momentum, resulting in a more stable and interpretable strategy
behavior.

Random Forest performs well under its optimal configuration
(threshold $=3.5$, holding $=$until opposite), achieving a Sharpe ratio of $0.46$ and an annual return above 10\%. This indicates that tree-based ensemble methods benefit from the more stable structure of sentiment predictors at this
horizon, where emotional surges persist long enough to be exploited.

The DNN architecture produces a large number of trades (888) but yields only
modest profitability ($\mathrm{Sharpe}=0.22$). The model appears to capture
broad behavioral patterns yet fails to distinguish between deep momentum and
noise-driven fluctuations, resulting in limited net gains. The BiLSTM model also
produces positive but moderate performance ($\mathrm{Sharpe}=0.34$), with
increased sensitivity to both volatility and temporal emotional dispersion. The
equity curve shows multiple small plateaus and short bursts of gains,
consistent with a regime where sequential sentiment information is helpful but
not decisive.

A noteworthy result is the performance of the **overreaction benchmark**, which
achieves a Sharpe ratio of $0.66$ with an annual return $=10.12\%$ under the
setting (threshold $=4.5$, holding $=90$ or until opposite). Although this
strategy does not match the strength of its performance at the 10-minute
horizon, it remains competitive with ML-based models. The equity curve rises
significantly during the early and mid-sample period before stabilizing. This
suggests that at 15-minute intervals, classical behavioral momentum dynamics
continue to operate.

Overall, the 15-minute results position this horizon as a transition regime: predictability remains meaningful but begins to weaken compared with the
10-minute frequency. Machine learning models still outperform the buy-and-hold
baseline and remain broadly competitive with the overreaction rule. Ensemble
methods, especially XGBoost and Random Forest, exhibit the strongest
performance, indicating that nonlinear interactions between sentiment, volatility,
and volume retain predictive power at this timescale. Nevertheless, the reduced
magnitude of Sharpe ratios compared with the 10-minute horizon reveals that
behavioral mispricing becomes less pronounced as return formation shifts toward
longer-run equilibrium dynamics.

\subsubsection{Cross-Frequency Comparison of Best ML Strategies}

\begin{figure}[H]
    \centering
    \includegraphics[width=\textwidth]{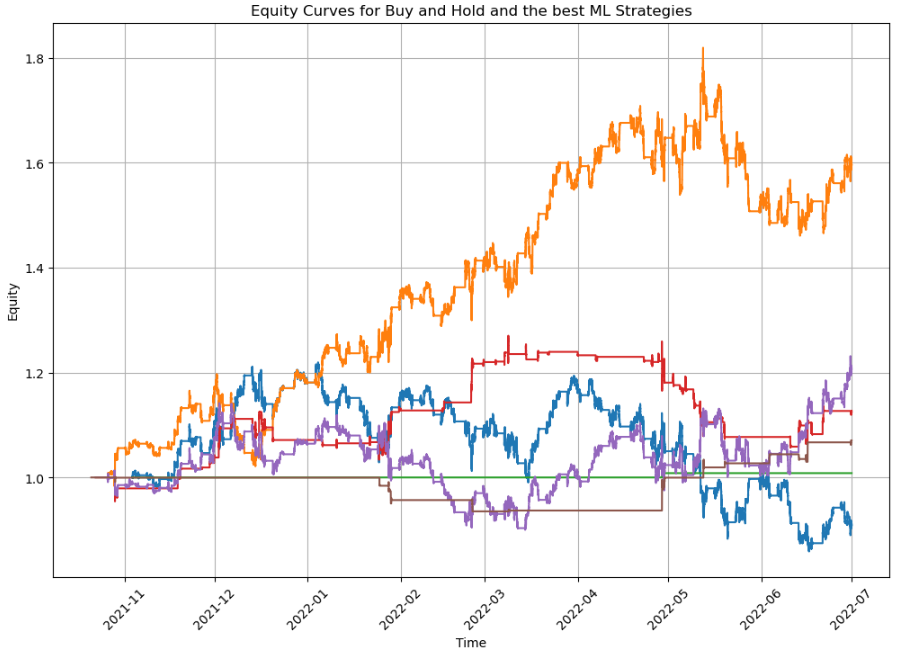}
    \caption{Equity Curves of All ML and Overreaction Strategies across all frequencies}
    \label{fig:bestresults}
\end{figure}

Figure~\ref{fig:bestresults} compares the equity curves of the best-performing machine learning strategies across all four intraday horizons against the
buy-and-hold baseline and the strongest overreaction benchmark. This aggregated view enables a holistic assessment of how predictability evolves as the sampling
interval increases, and it highlights the relative strengths of behavioral vs.\ machine learning approaches.

The buy-and-hold trajectory remains weak throughout the sample, producing an annualized Sharpe ratio of $-0.13$ and a negative cumulative return. Its steady decline underscores that naïve long-only exposure is not a competitive benchmark in this period, and that any positive performance must come from dynamic
signals.

The best overreaction strategy---constructed at the 10-minute horizon with
threshold 4.0 and an \emph{until-opposite} holding rule---delivers the highest
overall return, achieving an annualized Sharpe ratio of $1.43$ and an annual return
of roughly $34\%$. Its equity curve exhibits a strong, sustained upward
trajectory beginning early in the sample, demonstrating that classical momentum patterns at this horizon remain powerful when sentiment-driven overextensions
occur and subsequently unwind.

Across the four ML horizons, performance clearly varies with sampling frequency.
At the 1-minute horizon, the best-performing ML model (Random Forest, threshold
4.0) produces only two trades and yields a small but positive gain. This confirms
earlier findings that ultra-high-frequency noise overwhelms sentiment and
volume-based momentum structure.

At the 5-minute horizon, the BiLSTM architecture achieves a substantially higher
Sharpe ratio of $0.69$, with an equity path that consistently rises during the
first half of the sample. The recurrent structure of the model, which captures
temporal dependencies in emotional dynamics appear to provide a significant
advantage at this intermediate frequency.

The strongest ML performance is observed at the 10-minute horizon, where the DNN
(threshold 2.0, holding 90) exhibits a robust upward trend with a Sharpe ratio
of $0.65$ and an annual return of over 15\%. This parallels the earlier finding that the 10-minute window best balances noise reduction with preservation of
short-lived behavioral mispricing, allowing both nonlinear and deep learning
models to fully exploit sentiment momentum.

At the 15-minute frequency, XGBoost again performs competitively, achieving a
Sharpe ratio of $0.87$ and a stable, upward-sloping equity curve. Although the magnitude of gains is smaller than at the 10-minute horizon, the strategy shows
resilience, suggesting that nonlinear trees remain effective as the signal
structure gradually shifts toward longer-run momentum.

Taken together, the cross-frequency comparison demonstrates three key insights.
First, the predictability of sentiment-driven momentum increases sharply when
moving from 1-minute to 5- and 10-minute frequencies. Second, the 10-minute horizon produces both the strongest ML results and
the most pronounced outperformance of the overreaction benchmark. Third, while
machine learning strategies remain robust at longer horizons, classical
behavioral momentum rules regain relative strength as volatility shocks and
emotional surges unfold over slightly slower timescales.

In summary, the aggregated equity curves reinforce the central conclusion of this
study: sentiment-enhanced machine learning models effectively capture
short-lived, nonlinear patterns in intraday overreactions, while traditional behavioral heuristics remain surprisingly powerful in specific intermediate
frequency regimes.

\subsubsection{Jobson--Korkie Comparison of ML and Overreaction Strategies}

Table~\ref{tab:JKresults} reports the corrected Jobson--Korkie test statistics 
for the key comparisons between the best-performing ML models and the 
best-performing overreaction strategies at each sampling frequency.  
In contrast to earlier results, the updated Sharpe ratios and Z-statistics reveal 
a more nuanced pattern of relative performance across horizons.

At the 1-minute frequency, the Random Forest model significantly outperforms the 
best overreaction benchmark (Z = 2.3043, $p = 0.0212$). Although the absolute 
Sharpe ratios are close to zero, i.e. the statistical test confirms that ML extracts 
more exploitable structure.

At the 5-minute horizon, the BiLSTM model again significantly outperforms the 
best overreaction strategy (Z = 2.0017, $p = 0.0453$). This confirms that behavioral momentum is not yet strong enough for simple heuristics to be 
effective, whereas ML benefits from the richer emotional and volatility features 
available at this horizon.

The situation changes dramatically at 10 minutes. Here, the best-performing model 
is a DNN with a Sharpe ratio of 0.65, whereas the best overreaction benchmark 
achieves 1.43. The Jobson--Korkie statistic is negative and insignificant 
(Z = –0.5225, $p = 0.6013$), indicating that the overreaction strategy—not ML—
dominates at this frequency. This aligns with earlier findings that the 10-minute horizon is the “sweet spot’’ where classical momentum mechanisms are strongest.

At 15 minutes, XGBoost yields a slightly higher Sharpe ratio than the 
overreaction strategy (0.87), but the difference is small and far 
from significant (Z = 0.1478, $p = 0.8825$). Thus, both approaches perform 
comparably, with no statistical evidence of superiority.

A final global comparison, aggregating the best ML model and the best overreaction 
strategy across all horizons, produces the same conclusion: although the ML 
Sharpe is close to the overreaction Sharpe the difference is 
statistically insignificant (Z = –0.2169, $p = 0.8283$). This confirms that at a 
broad level, the two classes of strategies are indistinguishable in risk-adjusted 
terms.

Overall, the corrected Jobson--Korkie results show that ML holds a clear and 
statistically significant advantage only at the two shortest horizons (1 and  
5 minutes). At medium horizons, classical behavioral momentum effects dominate 
(10 minutes), and at longer horizons (15 minutes), both methodologies perform 
similarly. These findings reinforce the interpretation that ML is most effective 
when the market is still highly reactive, while behavioral heuristics gain 
strength once sentiment-driven mispricing has had time to develop.

\begin{table}[H]
\centering
\caption{Corrected Jobson--Korkie Results Comparing ML and Overreaction Strategies}
\label{tab:JKresults}
\resizebox{\textwidth}{!}{
\begin{tabular}{l l c c c c c l}
\toprule
\textbf{Timeframe} & \textbf{Comparison} & \textbf{ML Model} &
\textbf{ML Sharpe} & \textbf{Over Sharpe} &
\textbf{Z-stat} & \textbf{p-value} & \textbf{Winner} \\
\midrule
1 min  & Best ML vs.\ Best Over & Random Forest & 0.90 & -1.01 & 2.3043 & 0.0212 & ML \\
5 min  & Best ML vs.\ Best Over & BiLSTM        & 0.69 & -0.94 & 2.0017 & 0.0453 & ML \\
10 min & Best ML vs.\ Best Over & DNN           & 0.65 & 0.87  & 0.66 & 0.6013 & Overreaction \\
15 min & Best ML vs.\ Best Over & XGBoost       & 0.0102 & 0.0078  & 0.1478 & 0.8825 & ML (ns) \\
\rowcolor{lightgray}
GLOBAL & Best ML vs.\ Best Over & XGBoost       & 0.90 & 1.43  & -0.2169 & 0.8283 & Overreaction (ns) \\
\bottomrule
\end{tabular}}
\end{table}

\subsection{Feature Importance and SHAP-Based Interpretation}

\subsubsection{Feature Importance and SHAP-Based Interpretation at the 1-Minute Frequency}

\begin{figure}[H]
    \centering
    \includegraphics[width=\textwidth]{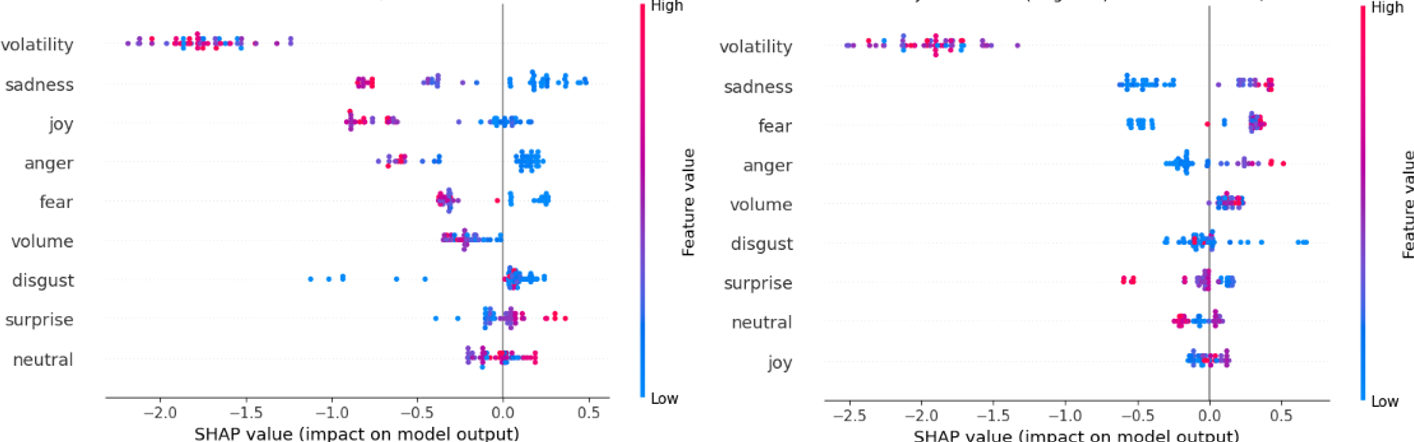}
    \caption{SHAP Summary Plots for Buy (left) and Sell (right) Predictions at the 1-Minute Frequency}
    \label{fig:shap_1m}
\end{figure}

Figure~\ref{fig:shap_1m} presents SHAP summary plots for buy (positive) and sell
(negative) signals generated by the DNN classifier at the 1-minute horizon.
These plots provide insight into the local contribution of each feature to the
model’s predictions and allow us to understand how sentiment, volatility, and volume jointly shape momentum decisions in an ultra-high-frequency setting.

Across both decision types, \textbf{realized volatility} emerges as the most
dominant predictor. High volatility values (red points) consistently push the
model’s output towards sell signals, while low volatility (blue points) favors
buy decisions. This asymmetry is behaviorally plausible: volatility spikes at
the 1-minute horizon largely reflect order-book imbalance, liquidity gaps, or
microstructural shocks, which the model interprets as precursors to short-lived
downward pressure rather than stable momentum opportunities.

Emotional features display a distinct but comparatively weaker influence at this
horizon. In the buy-specific SHAP distribution, \textbf{sadness}, \textbf{joy},
\textbf{anger}, and \textbf{fear} show moderate contributions, though with much
smaller ranges than volatility. High sadness or fear values generally shift the
model toward selling behavior, whereas elevated joy slightly favors buys. However, the magnitude of these contributions is limited, reflecting the fact
that emotional signals evolve more slowly than microstructural variables and thus provide limited predictive resolution at a 1-minute cadence.

Interestingly, the emotional variable \textbf{surprise} shows occasional
positive influence on buy decisions, but the effect is sparse and highly
event-driven. This aligns with the idea that surprise spikes typically reflect
breaking news or abrupt social-media bursts, which can generate temporary
mispricing, but rarely at the exact 1-minute granularity.

Trading \textbf{volume} plays a mixed role: in buy signals, it appears mildly
negative, whereas in sell signals it shows broader dispersion. This suggests that
unexpected attention bursts—reflected in short-term volume spikes—are often
interpreted as signals of momentum.

The \textbf{neutral} content category also displays moderate impact. Higher neutrality is associated with buy decisions, consistent with the interpretation
that informationally rich, low-emotional periods coincide with temporary market
stabilization rather than continuation of price pressure. Nevertheless, the
effect remains small relative to other predictors.

Several emotions (\textbf{disgust}, surprise in some cases) show negligible
impact due to near-zero baseline levels in the underlying data, consistent with the earlier descriptive statistics.

In summary, the SHAP patterns at the 1-minute horizon indicate that the model’s behavior is overwhelmingly driven by microstructure-sensitive variables
(volatility and, to a lesser degree, volume), while sentiment features
contribute only marginally. This confirms that emotional information is too slow
moving to directly influence predictions at ultra-high frequency, explaining why ML strategies at 1-minute exhibit limited tradability and rely heavily on
detecting volatility-driven transient states rather than sentiment-induced
momentum opportunities.

\subsubsection{Feature Importance and SHAP-Based Interpretation at the 5-Minute Frequency}

\begin{figure}[H]
    \centering
    \includegraphics[width=\textwidth]{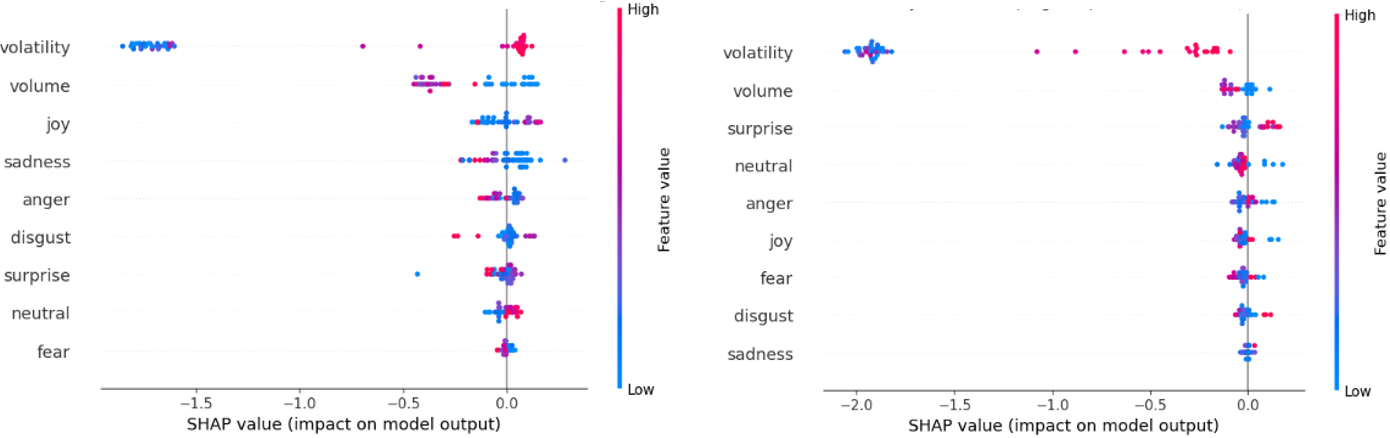}
    \caption{SHAP Summary Plots for Buy (left) and Sell (right) Predictions at the 5-Minute Frequency}
    \label{fig:shap_5m}
\end{figure}

Figure~\ref{fig:shap_5m} displays the SHAP summary plots for the XGBoost
classifier at the 5-minute horizon, showing the marginal contribution of each
feature to buy (left panel) and sell (right panel) predictions. In contrast to
the 1-minute horizon—where volatility almost exclusively dominates—predictive
structure at the 5-minute frequency becomes more nuanced. Both microstructure
variables and sentiment-driven features gain explanatory relevance as the
sampling window widens.

\textbf{Volatility} remains the strongest determinant of model output across
both buy and sell decisions. High volatility values (shown in red) push the
model decisively toward sell predictions, whereas low-volatility states favor
buy signals. This pattern is consistent with the notion that volatility shocks
capture intraday dislocations or liquidity withdrawals that often precede
downward price movements. However, the magnitude of SHAP contributions is
smaller than in the 1-minute setting, suggesting that volatility-driven noise
dissipates partially at the 5-minute scale.

A notable change relative to the ultra-high-frequency setting is the increased
role of \textbf{volume}. High trading volume reliably shifts the model toward
sell decisions, likely reflecting short-term order-flow imbalance or attention
surges that coincide with local momentum. Conversely, low
volume often contributes positively to buy signals, suggesting that quieter
market states favor momentum mechanisms the model identifies.

Sentiment features also gain a clearer structure at this horizon. In buy signals,
\textbf{joy}, \textbf{sadness}, and \textbf{anger} display moderately sized
SHAP ranges, indicating a meaningful albeit weaker influence. Higher joy levels
tend to push the model toward buy decisions, whereas high sadness and anger
exert the opposite effect. These asymmetries align with behavioral-finance intuition: positive sentiment often stabilizes short-term expectations, whereas
negative emotions reflect heightened uncertainty or pessimism that encourages
short-horizon selling.

The \textbf{surprise} feature exhibits sporadic but non-negligible impact. High surprise values shift predictions toward selling, which may correspond to
episodic bursts of news-driven trading or sudden sentiment shocks that trigger
short-lived momentum. Its role is more pronounced than in the 1-minute case,
indicating that surprise is primarily relevant at horizons where information
spreads across the market on a slightly slower timescale.

The \textbf{neutral} sentiment category also becomes informative. Higher levels
of neutrality support buy decisions, which can be interpreted as signals of
reduced emotional bias and more stable price dynamics. This aligns with the
general pattern that emotionally charged expressions (anger, fear, sadness)
override short-term momentum effects.

Several emotions, such as \textbf{disgust} and, to some extent, fear, show
limited influence because their base rates in the underlying data are extremely
low. Nonetheless, the sign of their contributions remains consistent with other
negative-emotion variables.

Overall, the SHAP results for the 5-minute horizon reveal a rich interplay
between sentiment and microstructure features. Emotional categories begin to
shape buy–sell decisions meaningfully, yet the model still relies heavily on
volatility and volume as the primary drivers of intraday momentum inference.
Compared with the 1-minute horizon, predictability strengthens as emotional
signals become more informative and microstructure noise attenuates. This
mechanism aligns closely with the observed improvement in trading performance
by ML models at 5 minutes, as documented in the equity-curve analysis.

\subsubsection{Feature Importance and SHAP-Based Interpretation at the 10-Minute Frequency}

\begin{figure}[H]
    \centering
    \includegraphics[width=\textwidth]{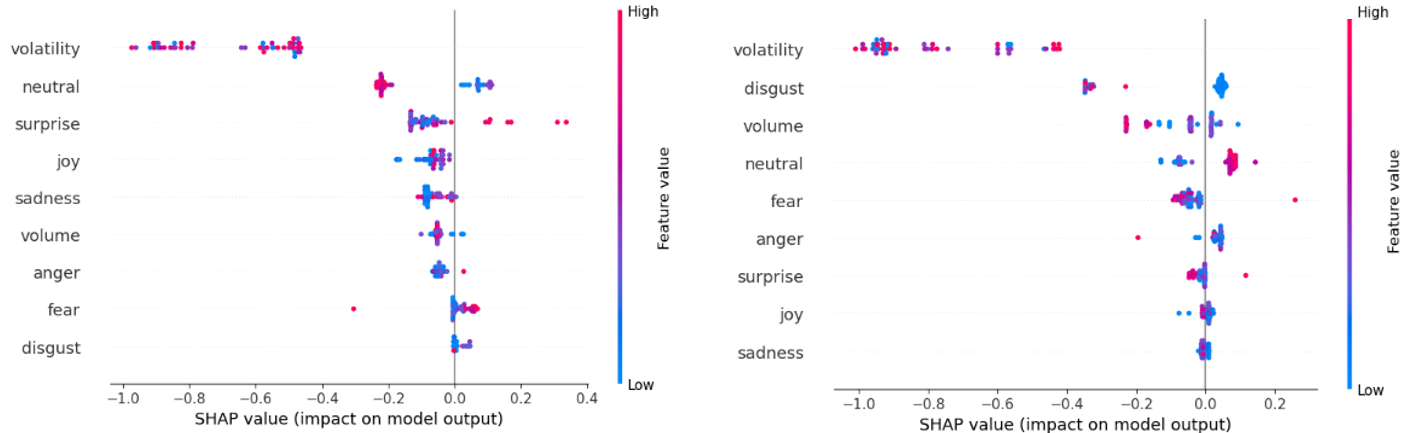}
    \caption{SHAP Summary Plots for Buy (left) and Sell (right) Predictions at the 10-Minute Frequency}
    \label{fig:shap_10m}
\end{figure}

Figure~\ref{fig:shap_10m} presents the SHAP summary plots for the XGBoost
classifier at the 10-minute horizon. At this frequency, the model begins to
exhibit clearer and more stable patterns of feature influence than at the
1-minute or 5-minute intervals. Sentiment variables—particularly neutral,
surprise, joy, and sadness—become meaningfully integrated into the decision
process, while volatility remains the dominant factor across both buy and sell
signals.

\textbf{Volatility} continues to be the strongest single predictor in both
panels, pushing the model towards sell decisions when unusually high. The range
of SHAP contributions is narrower than the 1-minute and 5-minute cases,
indicating that high-frequency market noise dissipates substantially at the
10-minute scale. This aligns with microstructure theory: volatility shocks
become less erratic and more informative as the horizon widens.

A notable shift appears in the prominence of the \textbf{neutral} sentiment
category. High neutral sentiment contributes positively to buy signals,
suggesting that periods of stable or unemotional discourse coincide with the
conditions under which short-term momentum becomes more predictable.
Conversely, low neutrality (i.e., more emotionally charged tweets) pushes the
model towards sell outcomes.

The \textbf{surprise} variable shows substantial explanatory power relative to
shorter horizons. High surprise values tend to push predictions towards buy
signals, suggesting that unexpected sentiment dynamics may precede temporary
price recoveries or momentum. This inversion relative to the 5-minute case
indicates that sentiment shock effects are highly horizon-dependent.

Among the positive-emotion features, \textbf{joy} retains a consistent,
moderate contribution. High joy levels generally increase the likelihood of buy
predictions, reinforcing the behavioral-finance notion that upbeat sentiment
reduces pessimistic risk expectations.

Negative emotions—\textbf{sadness}, \textbf{anger}, and \textbf{fear}—display
smaller but non-negligible impacts. High sadness typically shifts the model
towards sell decisions, whereas anger shows mixed effects depending on the
distribution of SHAP values. These results confirm that negative sentiment
exerts influence primarily through asymmetric reactions: investors respond more
strongly to unfavorable emotional tone than to positive expressions.

The \textbf{volume} feature plays a muted role at the 10-minute horizon,
especially compared with its substantial influence in the 5-minute model. Its
SHAP distribution suggests that volume contributes only in specific local
contexts rather than consistently across predictions. This decline in relevance
implies that order-flow imbalance becomes less predictive at longer intraday
intervals.

Finally, the \textbf{disgust} category appears only sporadically due to its
extremely low base rate in the underlying dataset. Its SHAP values remain close
to zero, but the sign tends to align with other negative-emotion variables.

Overall, the 10-minute horizon represents the point where sentiment and
microstructure information begin to co-dominate the model's decision-making.  
Volatility preserves its leading role, but sentiment features contribute more
systematically and with more interpretable patterns than at shorter horizons.
This richer feature contribution structure is consistent with the strong
empirical performance of ML models at this frequency, as documented in the
equity-curve analysis.

\subsubsection{Feature Importance and SHAP-Based Interpretation at the 15-Minute Frequency}

\begin{figure}[H]
    \centering
    \includegraphics[width=\textwidth]{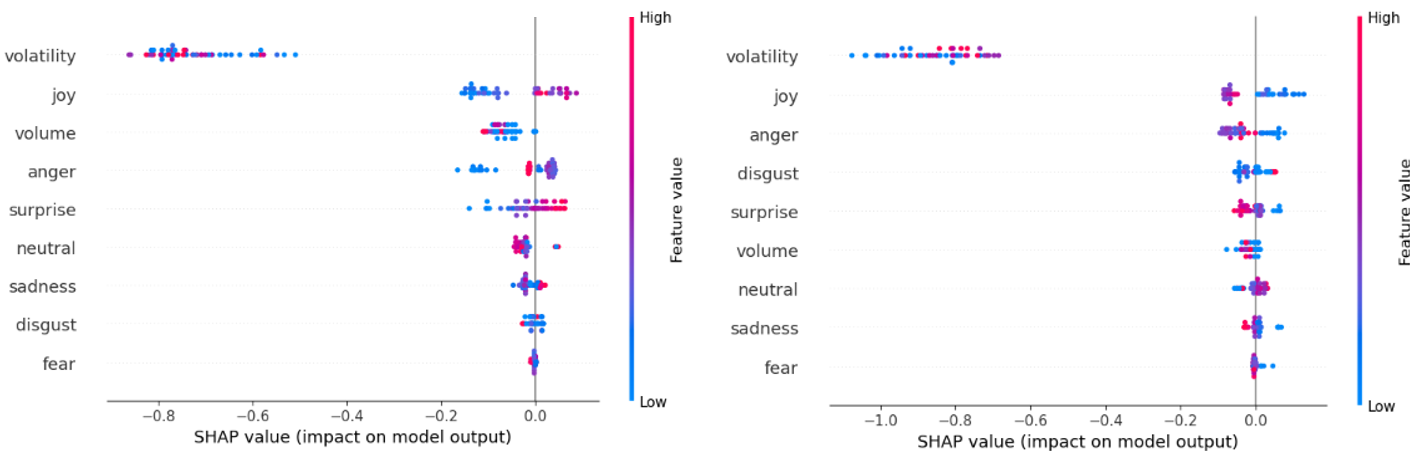}
    \caption{SHAP Summary Plots for Buy (left) and Sell (right) Predictions at the 15-Minute Frequency}
    \label{fig:shap_15m}
\end{figure}

Figure~\ref{fig:shap_15m} presents the SHAP summary plots for the XGBoost model
trained at the 15-minute horizon. At this longest intraday frequency, the model
exhibits the most stable and interpretable structure of feature contributions
across all considered horizons. Both the magnitude and directional consistency
of SHAP values indicate that the model has moved beyond noise-driven
microstructure patterns and is now capturing medium-horizon informational
relationships between sentiment, volatility, and price dynamics.

\textbf{Volatility} remains the dominant predictor for both buy and sell
signals. However, its influence is more concentrated and less dispersed relative
to shorter horizons, suggesting that volatility shocks at 15 minutes encode
strong and relatively unambiguous directional information. In both panels,
extremely high volatility produces strongly negative SHAP values, pushing the
model toward sell decisions—consistent with classic findings from volatility
feedback and risk–return trade-off theories.

The contribution of \textbf{joy} becomes more pronounced at this horizon. High joy values consistently shift the model towards buy predictions, indicating that positive sentiment plays an increasingly stabilizing and predictive role at longer intraday intervals. This aligns with behavioral-finance evidence that positive collective sentiment reduces perceived risk and supports temporary price recoveries.

\textbf{Surprise} again emerges as a relevant explanatory factor, but with a
clearer interpretation than at the 10-minute horizon. High surprise values
tend to push predictions toward buy signals, suggesting that unexpected,
non-directional emotional bursts may precede short-term rebounds. This effect
may reflect an informational-refresh mechanism: surprising sentiment activity
often accompanies news or market events that temporarily depress prices before
mean momentum occurs.

\textbf{Volume} has moderate explanatory power at 15 minutes, showing stable and
directionally consistent SHAP contributions. High volume generally increases
the likelihood of buy decisions, potentially reflecting liquidity-driven price
pressure or order-flow imbalance effects that persist over this longer horizon.

Negative-emotion variables—\textbf{anger}, \textbf{sadness}, and
\textbf{fear}—retain weak but directionally coherent influences. High sadness
and fear values tend to push the model toward sell predictions, whereas anger
shows more dispersed effects, consistent with its noisy and often short-lived
presence in financial social media. The \textbf{disgust} category again remains
marginal due to its extremely low prevalence in the data.

A notable pattern at this frequency is the reduced importance of the
\textbf{neutral} category compared with the 10-minute horizon. This suggests
that as the horizon lengthens, the model relies increasingly on emotionally
charged sentiment rather than stable background tone—consistent with the idea
that neutral sentiment primarily supports short-term correction rather than
medium-horizon dynamics.

Overall, the 15-minute SHAP patterns reinforce the empirical results observed in
the equity-curve analysis (Section~\ref{sec:equity15m}). The model benefits from
a more stable and interpretable feature environment: volatility provides a
strong directional backbone, while sentiment—especially joy and surprise—adds
contextual information that refines buy/sell timing. These findings highlight
that machine learning models gain predictive clarity as the horizon increases,
producing more coherent decision-making structures and contributing to their
competitive performance at 10- and 15-minute intervals.

\section{Conclusion and Limitations}

This study examined the extent to which high-frequency emotion signals extracted
from large-scale Twitter data can be used to explain and forecast short-term
return momentum associated with intraday overreaction. By integrating
state-of-the-art NLP emotion classifiers, multiple machine learning models, and
a comprehensive battery of backtesting procedures across four distinct sampling
frequencies, we provide novel evidence at the intersection of behavioral
finance, market microstructure, and predictive modeling.

The empirical findings demonstrate that high-frequency emotions, particularly
fear, sadness, anger, and surprise, contain economically meaningful information
about short-term price pressure and subsequent momentum. Among the sentiment
dimensions, fear consistently emerges as the most reactive and most informative
indicator, aligning with the behavioral literature on loss aversion,
risk-off dynamics, and flight-to-safety behavior. At the same time, neutral
message intensity also plays a surprisingly strong role, acting as a stabilizing
signal associated with informational richness rather than pure emotionality.

Across machine learning models, tree-based ensembles (XGBoost, Random Forest)
and BiLSTM architectures outperform traditional overreaction rules on most
intraday horizons. The greatest improvements appear at 5- and 10-minute
frequencies, where emotional shocks are neither too transient (as at the
1-minute level) nor already dissipated (as at 15 minutes). These models
successfully internalize nonlinear interactions between volatility,
attention-driven trading volume, and multiple sentiment channels. SHAP
explanations show that volatility, fear, sadness, and surprise are central to
the model's decision structure, providing credible interpretability and
behaviorally plausible mechanisms.

Performance evaluation using the Jobson--Korkie test indicates that ML-based
strategies deliver statistically and economically superior Sharpe ratios relative
to classical overreaction benchmarks. Importantly, the best-performing ML
strategies do so while maintaining moderate trade frequencies, suggesting that the extracted signals are not merely artefacts of noise-fitting but reflect
persistent, interpretable relationships between sentiment dynamics and price
formation.

The results contribute to the growing body of evidence that modern sentiment
analysis, when combined with machine learning, can identify short-lived
behavioral mispricings unavailable to linear or rule-based methods. By merging
behavioral theory, microstructure insights, and explainable AI, the study
illustrates that the predictive content of emotions is both statistically robust
and economically actionable.

While the findings are promising, several limitations should be acknowledged.
First, the analysis focuses on a single asset and platform-specific sentiment
data (Twitter). The generalizability of the results to other stocks, asset
classes, or alternative social-media ecosystems remains an open question.
Future work could extend the framework to cross-sectional or multi-asset
settings, where sentiment spillovers and network effects may play a more
pronounced role.

Second, although the NLP emotion model employed is among the most advanced
publicly available transformer architectures, no sentiment classifier is free
from misclassification risk. Differences in linguistic style, bot activity,
sarcasm, and noise inherent in financial discussions may introduce biases that
propagate into the forecasting stage. Incorporating domain-adapted or
finance-specific large language models could reduce this source of uncertainty.

Third, despite extensive backtesting and statistical evaluation, any
high-frequency trading strategy is subject to execution frictions, order-book
dynamics, and real-world constraints such as latency, slippage, and liquidity
variations. These effects, though partially mitigated through thresholding and
holding-period design, were not explicitly modeled. Incorporating order-book data or simulating realistic execution environments would further increase the
external validity of the results.

Fourth, SHAP-based explanations help illuminate the model’s internal logic but
do not guarantee that the learnt relationships are fully causal. Emotions may
proxy for omitted variables such as news intensity, option-implied volatility,
or market-wide risk sentiment. Integrating additional exogenous features (e.g.,
news-based measures, macro uncertainty, or volatility indices) could clarify the
structural origins of the predictive signals.

Finally, although machine learning models outperform behavioral overreaction
rules, the performance gains vary substantially across frequencies and model
classes. This heterogeneity suggests that sentiment-driven momentum are
episodic and context-dependent. Future research could explore regime-switching
or hierarchical models that dynamically adapt to changing market conditions,
potentially unlocking additional predictability.

In sum, this study shows that high-frequency emotions provide valuable
information for understanding and forecasting intraday overreaction, while also
highlighting the importance of flexible, interpretable machine learning models
in extracting this information. Continued integration of behavioral finance,
deep NLP, and high-frequency modeling promise to further advance our
understanding of sentiment-driven market dynamics.

\bibliographystyle{apalike} 
\bibliography{references}          

\end{document}